# MAC Address as a Key for Data Encryption

*Dr. Mohammed Abbas Fadhil Al-Husainy*
*Department of multimedia systems, faculty of science and information technology,*
Al-Zaytoonah University of Jordan.
Amman, Jordan
dralhusainy@yahoo.com , dralhusainy@gmail.com , alhusainy@zuj.edu.jo

*Abstract- In computer networking, the Media Access Control (MAC) address is a unique value associated with a network adapter. MAC addresses are also known as hardware addresses or physical addresses. TCP/IP and other mainstream networking architectures generally adopt the OSI model. MAC addresses function at the data link layer (layer 2 in the OSI model). They allow computers to uniquely identify themselves on a network at this relatively low level. In this paper, suggested data encryption technique is presented by using the MAC address as a key that is used to authenticate the receiver device like PC, mobile phone, laptop or any other devices that is connected to the network. This technique was tested on some data, visual and numerical measurements were used to check the strength and performance of the technique. The experiments showed that the suggested technique can be used easily to encrypt data that is transmitted through networks.*

## I. INTRODUCTION

Because of greater demand in digital signal transmission in recent time, the problem of illegal data access from unauthorized persons becomes need intelligent and quick solution. Accordingly, the data security has become a critical and imperative issue in multimedia data transmission applications. In order to protect valuable information from undesirable users or against illegal reproduction and modifications, various types of cryptographic/encryption schemes are needed. Cryptography offers efficient solutions to protect sensitive information in a large number of applications including personal data security, medical records, network security, internet security, diplomatic and military communications security, etc. through the processes of encryption/decryption.

Cryptography contains two basic processes: one process is when recognizable data, called plain data, is transformed into an unrecognizable form, called cipher data. To transform data in this way is called to encipher the data or encryption. The second process is when the cipher data is transformed back to the original plain data, this is called to decipher, or decrypting the data. To be able to determine if a user is allowed to access information a key is often used. Once a key has been used to encipher information, only someone who knows the correct key can decipher the encrypted data. The key is the foundation of most data encryptions algorithms today. A good encryption algorithm should still be secure even if the algorithm is known [1-5].

Encryption is the process of transforming the information to insure its security. With the huge growth of computer networks and the latest advances in digital technologies, a huge amount of digital data is being exchanged over various types of networks. It is often true that a large part of this information is either confidential or private. As a result, different security techniques have been used to provide the required protection [6].

MAC addresses are 12-digit hexadecimal numbers (48 bits in length). By convention, MAC addresses are usually written in one of the following two formats:

**MM:MM:MM:SS:SS:SS** or **MM-MM-MM-SS-SS-SS**

The first half of a MAC address contains the ID number of the adapter manufacturer. These IDs are regulated by an Internet standards body (see sidebar). The second half of a MAC address represents the serial number assigned to the adapter by the manufacturer. In the example: (00:A0:C9:14:C8:29), The prefix 00A0C9 indicates the manufacturer is Intel Corporation.MAC address spoofing is a synonym for taking over the identity of network interface controllers (NIC). Every single networking device is equipped with a globally unique hardware address called MAC address. The uniqueness of MAC addresses is essential in all phases of network communication because they map all upper-layer identifiers, e.g. IP addresses, to particular network interfaces[7].

Most difficult thing in security area is to assure the entity's identity. In many cases, string based pass phrases are used for this purpose. However, this kind of pass phrases is easily sniffed by a sophisticatedly architected key logging malware. Because of the reason, the highly security-sensitive services such as online finance and government issues are trying to restrict their operational environment in different ways. As an emphatic example of such approaches, Korean government has introduced the designated platform policy for online banking services where users are requested to use several limited number of PCs for renewing their public certificates. The problem in this approach is how to prove that the PC currently in use is one of the designated set of PCs. For this policy to be successful, it is essential to achieve the uniqueness of a PC platform to register it as a designated one [8].

The uniqueness of a hardware platform can be achieved by deriving platform-unique information from one or a combination of several hardware-dependent unique values. The Ethernet MAC address is considered the best one of such reasonable candidates as network IP address, serial numbers of hard disks, identifiers and mapping addresses of periphery devices and etc. because many people believe it is an un-modifiable and globally unique hardwired value. Although a MAC address needs to be unique only in a network segment, manufactures produce the Ethernet card with a pseudo globally-unique MAC address to eliminate the address conflicts when multiple cards are randomly deployed. In case of the designated platform solution [9], it utilizes the MAC address as a factor for constructing the platform-unique information. Therefore, this solution is somewhat for a kind of multi-factor authentication because the platform-unique information is used as an additional factor in proving both of the user identifier and the platform identifier. This approach can improve the security level of the services such as the online games, file repositories, financial transactions, etc. by restricting the locations (authenticated platforms) of the authenticated users. When a specific service is used, it registers the platform identifier to the management server. When the user tries to use this service back, the trials issued only on the platforms of which identifiers have registered are allowed and other requests are denied.

Practically, the MAC address is considered it should not be changed in an active service. Regarding the wireless gateway, it allows only the registered specific MAC addresses for network connections if configured especially in the case a mobile or vehicle machine makes an inquiry into a location in the dedicated network [10, 11].

With the advancements of multimedia and networks technologies, a vast number of digital images, video and other types of files now transmitted over Internet and through wireless networks for convenient accessing and sharing [5]. Multimedia security in general is provided by a method or a set of methods used to protect the multimedia content. These methods are heavily based on cryptography and they enable either communication security, or security against piracy (Digital Rights Management and watermarking), or both. Communication security of digital images and textual digital media can be accomplished by means of standard symmetric key cryptography. Such media can be treated as binary sequence and the whole data can be encrypted using a cryptosystem such as Advanced Encryption Standard (AES) or Data Encryption Standard (DES) [12]. In general, when the multimedia data is static (not a real-time streaming) it can treated as a regular binary data and the conventional encryption techniques can be used. Deciding upon what level of security is needed is harder than it looks. To identify an optimal security level, the cost of the multimedia information to be protected and the cost of the protection itself are to be compared carefully.

As a result, protection of digital images against illegal copying and distribution has become an important issue [5, 13, 14, 15].

There have been various data encryption techniques [16, 17, 18] on multimedia data proposed in the literature. Genetic Algorithms are among such techniques. The genetic algorithm is a search algorithm based on the mechanics of natural selection and natural genetics.

The genetic algorithm uses two reproduction operators: crossover and mutation. Reproduction give genetic algorithms most of their searching power. To apply a crossover operator, parents are paired together. There are several different types of crossover operators, and the types available depend on what representation is used for the individuals. The one-point crossover means that the parent individuals exchange a random prefix when creating the child individuals. The purpose of the mutation operator is to simulate the effect of transcription errors that can happen with a very low probability when a chromosome is mutated.

Only few genetic algorithms based encryption have been proposed. Kumar and Rajpal described encryption using the concept of the crossover operator and pseudorandom sequence generator by NLFFSR (Nonlinear Feed Forward Shift Register). The crossover point is decided by the pseudorandom sequence and the fully encrypted data they are able to achieve [19]. Kumar, Rajpal, and Tayal extended this work and used the concept of mutation after encryption. Encrypted data are further hidden inside the stego-image [20].

Husainy proposed Image Encryption using Genetic Algorithm-based Image Encryption using mutation and crossover concept [21].

A. Tragha et al., describe a new symmetrical block ciphering system named ICIGA (Improved Cryptography Inspired by Genetic Algorithms) which generates a session key in a random process. The block sizes and the key lengths are variable and can be fixed by the user at the beginning of ciphering. ICIGA is an enhancement of the system (GIC) "Genetic algorithms Inspired Cryptography" [22].

A different technique for secure and efficient data encryption has been presented in this paper. This technique employs the MAC address of the receiver as a key to encrypt data, crossover and mutation operations of the Genetic Algorithm (GA) are using here with the MAC address to produce a strong encrypted data that have a good immunity against the attackers.

II. MATERIALS AND METHODS

Whenever the sender want send a safe data to the receiver, and after establishing the communication session. The MAC address of the receiver's device is read by the encryption technique to use it as a key to encrypt the data. The six parts of the MAC address will be formed to represent a vector (chromosome) of 6 bytes (genes). For example: (00:A0:C9:14:C8:29) is represent (in decimal) as:

| 0 | 160 | 201 | 20 | 200 | 41 |
|---|---|---|---|---|---|

Digital data file will be treated as a set of N bytes. The data is read and splitting it into a set of (N/6) vectors (chromosomes) of 6 bytes (same length as the MAC address above). For example, if the source data file has 24 bytes as:

| 2 | 10 | 7 | 15 | 32 | 19 |
|---|---|---|---|---|---|
| 9 | 64 | 71 | 3 | 15 | 23 |
| 1 | 12 | 34 | 18 | 5 | 25 |
| 30 | 11 | 3 | 16 | 27 | 8 |

Then these bytes are represented as:

Vector #

| 1 | 2 | 10 | 7 | 15 | 32 | 19 |
|---|---|---|---|---|---|---|
| 2 | 9 | 64 | 71 | 3 | 15 | 23 |
| 3 | 1 | 12 | 34 | 18 | 5 | 25 |
| 4 | 30 | 11 | 3 | 16 | 27 | 8 |

Now, the technique performs three main operations on the data vectors above:

1. Crossover or transposition the gene order in each vector. This is done by using a pseudo random number generation algorithm with different seed (initial) value for each vector (the vector number in this work). After doing this operation the data vectors become as follow:

Vector #

| 1 | 32 | 19 | 2 | 7 | 15 | 10 |
|---|---|---|---|---|---|---|
| 2 | 23 | 71 | 64 | 9 | 3 | 15 |
| 3 | 25 | 18 | 34 | 5 | 1 | 12 |
| 4 | 27 | 3 | 30 | 8 | 11 | 16 |

2. Mutation or substitution the value of each gene in each vector. This is done by apply an eXclusive-OR (XOR) Boolean operation between the MAC address vector and each of the data vector. After finish this operation the data vectors become as follow:

Vector #

| 1 | 32 | 179 | 203 | 19 | 199 | 35 |
|---|---|---|---|---|---|---|
| 2 | 23 | 231 | 137 | 29 | 203 | 38 |
| 3 | 25 | 178 | 255 | 17 | 201 | 37 |
| 4 | 27 | 163 | 215 | 28 | 195 | 57 |

3. Re-sequence or reorder the sequence of the vectors itself. To make extra distortion in the encrypted data vector, the technique reorders the sequence of the data vectors randomly to be as follow:

Vector #

| 3 | 25 | 178 | 255 | 17 | 201 | 37 |
|---|---|---|---|---|---|---|
| 4 | 27 | 163 | 215 | 28 | 195 | 57 |
| 1 | 32 | 179 | 203 | 19 | 199 | 35 |
| 2 | 23 | 231 | 137 | 29 | 203 | 38 |

When the three main operations are completed, the technique produces the encrypted data file which is being as follow:

| 25 | 178 | 255 | 17 | 201 | 37 |
|---|---|---|---|---|---|
| 27 | 163 | 215 | 28 | 195 | 57 |
| 32 | 179 | 203 | 19 | 199 | 35 |
| 23 | 231 | 137 | 29 | 203 | 38 |

To ensure that the encryption technique really will happen enough distortion in the source data, the measurement of Signal to Noise Ratio (SNR) can be used here. The SNR is calculated by using the following formula, where S and E represent the source and the encrypted image respectively:

$$SNR_{db} = \frac{\sum_{i=1}^{width} \sum_{j=1}^{hieght} (E_{ij})^2}{\sum_{i=1}^{width} \sum_{j=1}^{hieght} (E_{ij} - S_{ij})^2} \quad (1)$$

For the numerical example above, SNR between the source and the encrypted data ≅ (1.200 db). This ratio is enough to make a good protection to the source data against the attackers.

### III. RESULTS AND DISCUSSION

To give reader an ability to note the performance of the suggested technique, an experiment is implemented on an image of type (.bmp) to see strength of the encryption technique visually. The required programming codes to implement the proposed method are written using JAVA programming language.

Key space analysis, key sensitivity analysis, statistical analysis and Signal to Noise Ratio (SNR) are some of the security tests that are recommended to be used for testing the performance, strength and immunity of encryption methods.

#### A. Key space analysis

In any effective encryption system, the key space should be large enough to make brute-force attack infeasible. The secret key space (MAC address) in the suggested technique is (6bytes = 48bits), this means that the encryption system has relatively enough number of bits in the secret key. In this work, we note that the bits in the key are restricted by the MAC address and they cannot be increased or decreased.

#### B. Key sensitivity

To evaluate the key sensitivity feature of the proposed technique, a one bit change is made the secret key (MAC address) and then used it to decrypt the encrypted image. The decrypted image with the wrong key is completely different when it is compared with the decrypted image by using the correct key as shown in Fig. 1. It is the conclusion that the proposed encryption technique is highly sensitive to the key, even an almost perfect guess of the key does not reveal any information about the plain image/data.

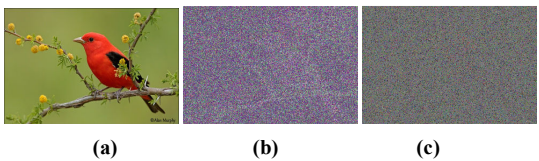

(a)      (b)      (c)

**Figure 1: (a) Source image (b) Encrypted image (c) Decrypted image with wrong key**

#### C. Statistical analysis

Statistical attack is a commonly used method in cryptanalysis and hence an effective encryption system should be robust against any statistical attack. Calculating the histogram and the correlation between the neighbors pixels in the source and in the encrypted image are the statistical analysis to prove the strong of the proposed encryption system against any statistical attack.

Fig. 2 shows the histograms of the source image in Figure 1 and its encrypted image respectively. It's clear from Fig. 2 that the histogram of the encrypted image is completely different from the histogram of the source image and does not provide any useful information to employ statistical attack.

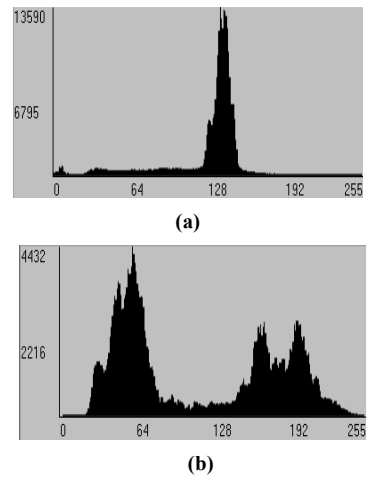

(a)

(b)

**Figure 2: (a) Histogram of the source image in Fig. 1(a)**

**(b) Histogram of its encrypted image**

The correlation coefficient r is calculating by using the following formula:

$$r = \frac{\sum_{i=1}^{N}(x_i - \overline{x})(y_i - \overline{y})}{\sqrt{\sum_{i=1}^{N}(x_i - \overline{x})^2 \times \sum_{i=1}^{N}(y_i - \overline{y})^2}} \quad (2)$$

Where N is the number of pixel pairs,

$$\overline{x} = \frac{1}{N}\sum_{i=1}^{N} x_i \quad (3)$$

And

$$\overline{y} = \frac{1}{N}\sum_{i=1}^{N} y_i \quad (4)$$

The correlation coefficient for horizontal neighbor pixels of the source image in Fig. 1 is r=0.100231 while r=0.005561 for its encrypted image. It is clear from these two different values of the correlation coefficient that the strong correlation between neighbor pixels in source image is greatly reduced in the encrypted image. The results of the correlation coefficient for vertical and diagonal neighbor pixels are similar to the horizontal neighbor pixels. The Signal to Noise Ratio (SNR) that is calculated for the above encrypted image in Fig. 1(b) is SNR=4.04401.

Some other images and data files were tested; the same behavior in the suggested encryption technique has been recorded.

From the above test, we note the following points:

The low value of SNR refer to that there is much distortion in the encrypted image. This means that the encrypted image has good immunity against the Human Visual System (HVS) attack.

The value of the correlation coefficient of the encrypted image is reducing heavily. And it is minimized greatly when comparing its value with the value of the correlation coefficient of the source image.

## IV. CONCLUSIONS

In this paper, a technique for data encryption has been presented which employ the MAC address of the receiver device to use it as a key for encryption. This technique made a good immunity for the data that is transmitted through networks. The visual and analytical tests showed that the suggested technique is useful to use in the field of image/data encryption effectively in networks.

AUTHOR

**Mohammed Abbas Fadhil Al-Husainy** received the M.Sc. and Ph.D. degrees in 1996 and 2002, respectively. From 1997 to 2002, he was a lecturer in the Department of Computer Science, Al-Hadba University of Mosul. Since 2002 he has been an associate professor in the Departments: Computer Science and Multimedia Systems, Faculty of Science and Information Technology, Al-Zaytoonah University of Jordan. He lectures in the areas of microprocessors, data structures, algorithm design and analysis, digital design systems, operating systems, cryptography, computer organization, programming languages. His research interests are in the broad field of algorithm design, including multi-media data processing, scheduling algorithms, and cryptography algorithms.